\documentclass{aa}

\usepackage{graphicx}
\usepackage{txfonts}

\begin{document}

\title{GCIRS34W: An irregular variable in the Galactic Centre \thanks{Based on observations at the New Technology Telescope (NTT) and the Very Large Telescope (VLT) of the European Southern Observatory (ESO), Chile, the 2.2-m-telescope of the Max-Planck-Gesellschaft (MPG) in La Silla, Chile, and the Gemini North telescope of the Gemini Observatory on Mauna Kea, Hawaii.}}

\author{S. Trippe \inst{1} \and F. Martins \inst{1} \and T. Ott \inst{1} \and T. Paumard \inst{1} \and R. Abuter \inst{1} \and F. Eisenhauer \inst{1} \and S. Gillessen \inst{1} \and R. Genzel \inst{1,2} \and A. Eckart \inst{3} \and R. Sch\"odel \inst{3}}

\offprints{S. Trippe, \email{trippe@mpe.mpg.de}}

\institute{Max-Planck-Institut f\"ur extraterrestrische Physik, Giessenbachstrasse, Postfach 1312, D-85741 Garching, Germany \and Department of Physics, University of California, Berkeley, USA \and 1. Physikalisches Institut, Universit\"at zu K\"oln, Z\"ulpicher Strasse 77, D-50937 K\"oln, Germany}

\date{Received April 1, 2005 / Accepted October 17, 2005 }

\abstract{We report the results of time-resolved photometric and spectroscopic near-infrared observations of the Ofpe/WN9 star and LBV candidate GCIRS34W in the Galactic Centre star cluster. Diffraction limited resolution photometric observations obtained in $H$ and $K$ bands show a strong, non-periodic variability on time scales from months to years in both bands accompanied by variations of the stellar colour. Three $K$ band spectra obtained in 1996, 2003 and 2004 with integral field spectrometers are identical within their accuracies and exclude significant spectroscopic variability. The most probable explanation of the stellar photometric variability is obscuration by circumstellar material ejected by the star. The approximated position of GCIRS34W in a HR diagram is located between O supergiants and LBVs, suggesting that this star is a transitional object between these two phases of stellar evolution.

}

\maketitle

\section{Introduction}

The class of Ofpe/WN9 stars has received attention in the
last decade. Initially defined as stars with spectroscopic characteristics of both the
most extreme O supergiants and late WN stars (Walborn \cite{Walborn}), these objects may be
in an intermediate evolutionary state between these two types of stars
(Bohannan \& Walborn \cite{Bohannan89}, Nota et al. \cite{Nota}, Pasquali et al. \cite{Pasquali}). Since the observation
by Stahl (\cite{Stahl83}) of the Ofpe/WN9 star R127 turning to a Luminous Blue Variable (LBV), a link between these two
classes of objects is suspected. Stahl (\cite{Stahl86}) also showed that the LBV AG Car displays
a typical Ofpe/WN9 star in its minimum phase and Crowther et al. (\cite{Crowther}) associated a
few Ofpe/WN9 stars with dormant LBVs from comparison of their chemical composition and
physical parameters.

LBVs are well known for their strong, episodic variability (Conti \cite{Conti}).
These stars are usually very luminous ($L \simeq 10^{6} L_{\odot}$). They can experience "great eruptions" with $V$ increasing by more than 2 mag (e.g. $\eta $ Car, P Cyg) or they
can stay in a relatively quiescent phase with low $V$ band variation ($\simeq$ 0.5 mag) and
suddenly erupt with brightening of 1-2 mag (see Humphreys \& Davidson \cite{Humphreys94} for a
review).
All mechanisms developed to explain the behaviour of LBVs (Glatzel \& Kiriakidis (\cite{Glatzel}), Dorfi \& Gautschy (\cite{Dorfi}), Humphreys \&
Davidson (\cite{Humphreys84}), Lamers \& Fitzpatrick (\cite{Lamers88}), Nugis \& Lamers (\cite{Nugis})) are associated with strong mass outflows (with $\dot{M}$
as high as $0.1 M_{\odot}$/yr in case of $\eta$ Car, Massey \cite{Massey}) which is confirmed by the
detection of nebulae around LBVs (Voors et al. \cite{Voors}). Hence spectroscopic variability
due to both changes in $T_{\rm eff}$ and $\dot{M}$ is also common in LBVs.

In this article we present the results of photometric and spectroscopic long-term observations of the star GCIRS34W (hereafter referred to as IRS34W) which is part of the star cluster surrounding the supermassive black hole in the centre of our Milky Way. This star cluster, located in a distance of 7.6 kpc (Eisenhauer et al. \cite{Eisenhauer05}), is a system of dynamically relaxed old stars with an admixture of young, hot, non-relaxed stars; some of the young, hot stars show strong \ion{He}{I} emission (Krabbe et al. \cite{Krabbe91}). Most of the young and massive stars appear to reside in two thin rotating disks centered on the Galactic Centre supermassive black hole Sgr A* (Genzel et al. \cite{Genzel03a}, Paumard et al. \cite{Paumard05}). Among these objects, around 30 have been identified as Wolf-Rayet stars and LBV candidates (Paumard et al. \cite{Paumard04}). Another group of early- and late-type B main sequence stars can be found in the immediate vicinity (few tens of light days) of Sgr A* (Genzel et al. \cite{Genzel03a}, Eisenhauer et al. \cite{Eisenhauer05}), leading to the so-called "paradox of youth", i.e. the question, how these massive and young stars managed to reside in an area so close to the black hole (Ghez et al. \cite{Ghez}).

IRS34W is one of the young \ion{He}{I} stars and was already identified as a Ofpe/WN9 star in earlier examinations (Krabbe et al. \cite{Krabbe95}, Najarro et al. \cite{Najarro}); additionally, IRS34W has been classified as a LBV candidate by Paumard et al. (\cite{Paumard01}). Therefore this star is of great interest for probing the understanding of star formation and evolution in the Galactic Centre.

In this article we describe the observations and data reduction in Sect. 2, present and discuss the reults in Sect. 3 and summarize our conclusions in Sect. 4.

\section{Observations}

\subsection{Photometry}

Photometric observations of IRS34W have been obtained regularly since 1992.

From 1992 to 2002 we used the camera SHARP I (Hofmann et al. \cite{Hofmann}) at the 3.5-m-NTT of ESO in La Silla, Chile, in speckle imaging mode to obtain diffraction limited resolution $K$ band images. All images were sky-subtracted, bad-pixel- and flat-field-corrected and deconvolved with a Lucy-Richardson-algorithm (Lucy \cite{Lucy}, Richardson \cite{Richardson}), the spatial resolution was 130 milli-arcseconds (mas). On the processed images we applied aperture photometry, using a set of four stars located in the observed field as calibrators. Additionally in 1992, 1997 and 1998 we obtained $H$ band images with a spatial resolution of 100 mas. These images were processed (except deconvolution) and analyzed like the $K$ images.

Since 2002 we used the detector system NAOS/CONICA (NACO for short) consisting of the adaptive optics system NAOS (Rousset et al. \cite{Rousset}) and the NIR camera CONICA (Hartung et al. \cite{Hartung}) at the 8.2-m-UT4 (Yepun) of the ESO-VLT on Cerro Paranal, Chile. We obtained AO-corrected diffraction limited resolution images in $H$ (spatial resolution 40 mas) and $K$ bands (spatial resolution 55 mas). After sky subtraction, bad-pixel- and flat-field-correction all images were deconvolved with a Wiener filtering algorithm (Ott et al. \cite{Ott99}). On the deconvolved images we applied aperture photometry, here using an ensemble of 12 stars in the field of view as reference sources.

Additional to our observations we included images from the Galactic Center Demonstration Science Data Set obtained in 2000 with the 8-m-telescope Gemini North on Mauna Kea, Hawaii, using the AO system Hokupa'a in combination with the NIR camera Quirc. These images were processed by the Gemini team and released to be used freely. We analyzed one $H$ and one $K$ band image obtained in July 2000 by applying aperture photometry calibrated with a set of three stars.

\begin{figure}[t]
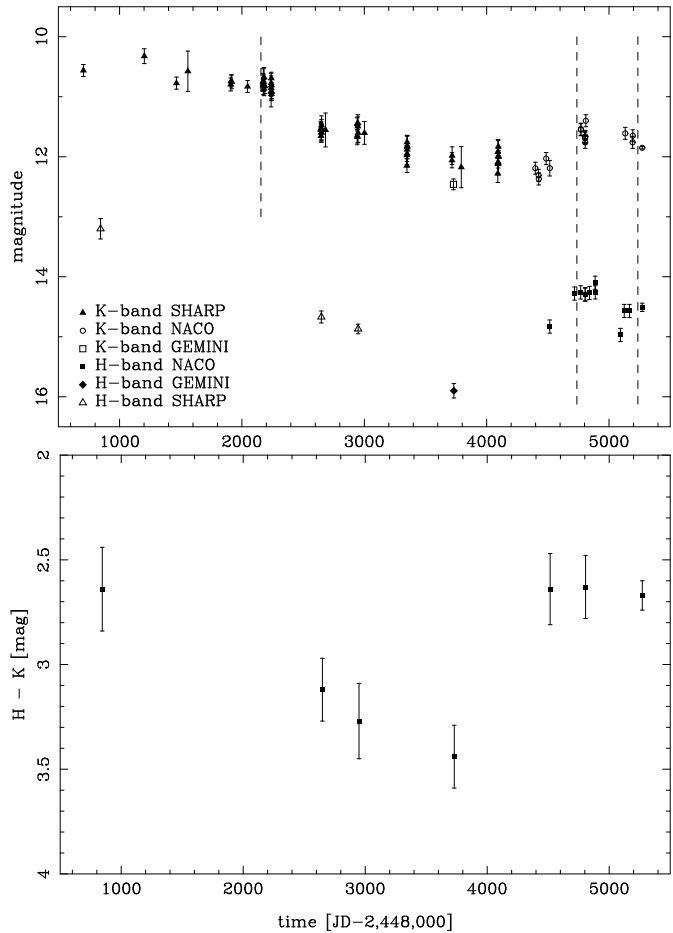

\resizebox{\hsize}{!}{\includegraphics[angle=-90]{trippe01.eps}} \\
\resizebox{\hsize}{!}{\includegraphics[angle=-90]{trippe04.eps}}
\caption{{\it Top panel:} H and K band lightcurves of IRS34W. The time is given as reduced Julian Date, the first data point was taken in March 1992, the last in September 2004. The vertical dashed lines indicate the times when the three spectra were obtained in 1996, 2003 and 2004. Strong non-periodic variabillity on time scales from months to years is clearly visible in both light curves. {\it Bottom panel:} H-K derived from the lightcurves. The comparison to the lightcurves shows that the colour gets redder when the magnitudes decrease, and bluer when the magnitudes increase.}
\label{photo}
\end{figure}

\subsection{Spectroscopy}

For a time-resolved spectroscopic analysis of IRS34W we compared three spectra obtained in 1996, 2003 and 2004.

\begin{figure}[t]
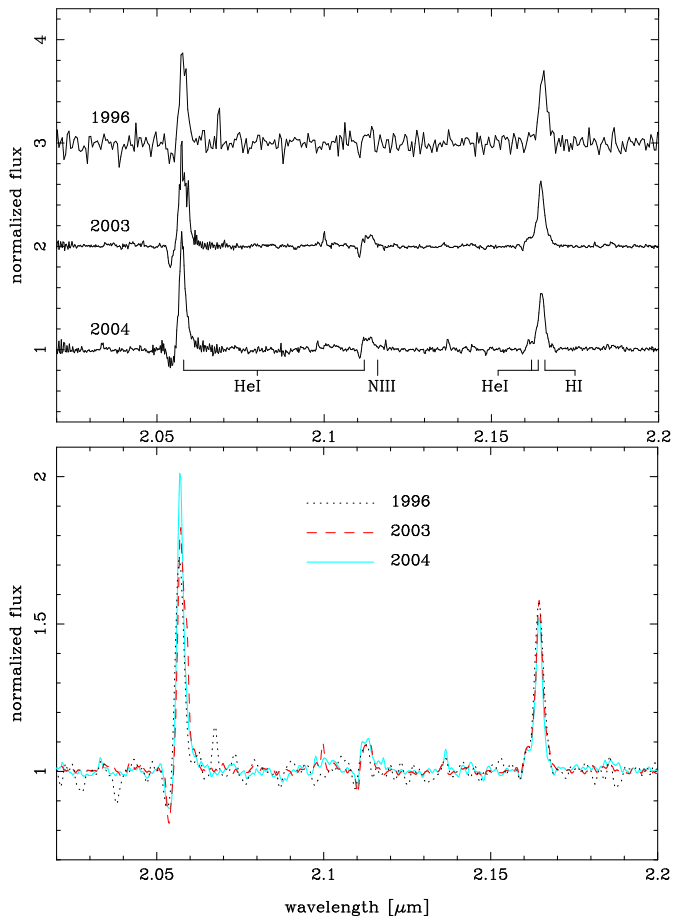

\resizebox{\hsize}{!}{\includegraphics[angle=-90]{trippe02.eps}} \\
\resizebox{\hsize}{!}{\includegraphics[angle=-90]{trippe05.eps}}
\caption{{\it Top panel:} Normalized $K$ band spectra of IRS34W obtained in March 1996 with 3D, in April 2003 with SPIFFI and in August 2004 with SINFONI (spectra shifted along the flux axis). {\it Bottom panel:} The same spectra, smoothed and overplotted. Obviously, the spectra are identical within their typical accuracies.}
\label{spectra}
\end{figure}

The first spectrum was obtained in March 1996 using the integral field spectrometer 3D (Weitzel et al. \cite{Weitzel}) at the 2.2-m-MPG-telescope in La Silla, Chile. The data output is structured as data cubes with two spatial axes, each 16 pixels long, and one spectral axis of 256 pixels length. So 3D provides seeing limited 16$\times$16 pixel images with a spectrum for each image pixel. The March 1996 observations covered the wavelength interval from 1.97 to 2.21$\mu$m (the short wavelength half of $K$ band) with a spectral resolution of $R = 1500$. The seeing was around 0.6'', the spatial pixelscale was 300 mas/pixel. After sky subtraction, bad-pixel- and flat-field-correction the spectrum was calibrated spectrally with a neon lamp. Atmospheric features were removed by dividing by the normalized spectrum of a calibration star, only in the range of low atmospheric transmission and high noise below 2.02$\mu$m the correction was incomplete.

The second spectrum was obtained in April 2003 with the integral field spectrometer SPIFFI (Eisenhauer et al. \cite{Eisenhauer03a}, \cite{Eisenhauer03b}) at the VLT-UT2 (Kueyen). SPIFFI produced seeing limited data cubes with 32 pixels in each spatial axis and 1024 pixels in the spectral axis. The data were corrected and calibrated (sky, bad pixels, flat field, atmosphere, wavelength calibration) like the 3D spectrum. Seeing was around 0.3'', the spatial pixelscale was 100 mas/pixel. The final spectrum covered the complete $K$ band from 1.95 to 2.45$\mu$m with a spectral resolution of $R = 3600$.

The third spectrum was obtained in August 2004 using SINFONI, a combination of SPIFFI and the adaptive optics (AO) system MACAO (Bonnet et al. \cite{Bonnet03}, \cite{Bonnet04}) at VLT-UT4. As SPIFFI was equipped with a new 2K$\times$2K detector in early 2004, the data cubes know had a dimension of 64$\times$32 pixels in the two spatial dimensions and 2048 pixels along the spectral axis. The data were processed and reduced like the earlier SPIFFI data. The pixel scale of the images was 100$\times$50 mas/pixel, meaning that each image pixel is rectangular. Observations were obtained at a seeing of around 0.8'', the spatial resolution was improved to 0.3'' by the AO system. The spectrum covered the $K$ band from 1.95 to 2.45$\mu$m with a spectral resolution of $R = 4500$.

We corrected each spectrum for background flux by subtracting a spectrum obtained in an empty field 0.9'' north of IRS34W. To assure the comparability of the spectra and to estimate the influence of contamination by the neighbouring stars (especially by IRS34E located 0.4'' north east of IRS34W), we obtained the 2003 SPIFFI spectrum twice: once on the original data cube, and once on a data cube in which (1) the spatial resolution (0.3'') was reduced to 3D resolution (0.6'') by convolving the spatial images with a gaussian, (2) the image pixels (pixel scale 100 mas/pixel) were combined to pixels of the size of 3D pixels (300 mas/pixel) and (3) the spectral resolution ($R = 3600$) was reduced to that of 3D ($R = 1500$) by convolving the spectral axis with a gaussian.

\section{Results and discussion}

The $H$ and $K$ band light curves and colour of IRS34W obtained from the stellar photometry are presented in Fig. \ref{photo}. The magnitudes used in this discussion are not corrected for the (not well-known) extinction by interstellar matter, the given times are days after JD 2,448,000.

Both lightcurves show variations in magnitude on timescales between few 100 and around 1500 days, a selection of representative data points (one value per calendar year) is presented in Table \ref{mags}. Especially the $K$ lightcurve, showing an overall variation of $\Delta K \simeq 1.5$, does not show any periodicity or regularity, including periods of nearly constant magnitude as well as steplike changes in brightness.

The stellar colour $H-K$ shows variations which are clearly correlated to the lightcurves. From day 845 to day 3731 $H-K$ gets redder by about 0.8, whereas in the same time $K$ decreases by 1.5. Within days 3731 and 4516 $K$ increases again by around 0.4 and the colour gets bluer by roughly 0.8. We do not observe, within the errors, a significant variation in the time from day 4516 to day 5271, while in the same time span $H$ and $K$ both vary by around 0.5 magnitudes.

Variations on small timescales (few days) are not observed. Although the SHARP $K$ lightcurve shows groups of data points not resolved in time (they were obtained in observing runs lasting 4--10 days) which appear to be somewhat scattered, these values are identical within their errors. One should note that the Gemini $K$ point at day 3728 appears to be slightly offset compared to the SHARP points; but taking into account the error bars and the fact that these points were extracted from two different, seperately processed data sets, this is no significant indication towards short time variability either.

\begin{table}[t]
\caption{Selected $H$ and $K$ magnitudes (one data point per calendar year) and colours of IRS34W, the values are not corrected for extinction. The time is given in days after JD 2,448,000. A months- to years-time-scale variability can be seen in $H$ and $K$, including variations of the stellar colour.}
\label{mags}
\centering
\begin{tabular}{l c c c}
\hline\hline
time & $H$ & $K$ & $H-K$ \\
\hline
705  & & 10.56$\pm$0.10 & \\
     & & & 2.64$\pm0.20^{\mathrm{a}}$ \\
845  & 13.20$\pm$0.17 & & \\
1203 & & 10.33$\pm$0.12 & \\
1465 & & 10.77$\pm$0.10 & \\
1917 & & 10.75$\pm$0.12 & \\
2241 & & 10.81$\pm$0.11 & \\
2649 & 14.67$\pm$0.10 & 11.55$\pm$0.11 & 3.12$\pm$0.15 \\
2949 & 14.87$\pm$0.08 & 11.60$\pm$0.16 & 3.27$\pm$0.18 \\
3350 & & 11.95$\pm$0.13 & \\
3728 & & 12.46$\pm$0.09 & \\
     & & & 3.44$\pm0.15^{\mathrm{b}}$ \\
3731 & 15.90$\pm$0.12 & & \\
4094 & & 12.09$\pm$0.10 & \\
4516 & 14.83$\pm$0.11 & 12.19$\pm$0.13 & 2.64$\pm$0.17 \\
4806 & 14.29$\pm$0.11 & 11.66$\pm$0.10 & 2.63$\pm$0.15 \\
5271 & 14.51$\pm$0.07 & 11.85$\pm$0.02 & 2.67$\pm$0.07 \\
\hline

\end{tabular}

\begin{list}{}{}
\item[$^{\mathrm{a}}$] This colour is computed from the magnitudes at days 705 and 845.
\item[$^{\mathrm{b}}$] This value is computed from the magnitudes at days 3728 and 3731. Both $H$ and $K$ were extracted from the Gemini North data set.
\end{list}

\end{table}

The normalized spectra of IRS34W obtained in 1996, 2003 and 2004 are presented in Fig. \ref{spectra}, all wavelengths used below are given in microns. All spectra show as prominent features the lines \ion{He}{I} ($\lambda$ 2.058), \ion{He}{I} ($\lambda$ 2.112), \ion{N}{III} ($\lambda$ 2.116), \ion{He}{I} ($\lambda$ 2.162), \ion{He}{I} ($\lambda$ 2.164) and \ion{H}{I} ($\lambda$ 2.166) in emission. Due to line blending the lines \ion{He}{I} ($\lambda$ 2.112) and \ion{N}{III} ($\lambda$ 2.116) on the one hand and \ion{He}{I} ($\lambda$ 2.162), \ion{He}{I} ($\lambda$ 2.164) and \ion{H}{I} ($\lambda$ 2.166) on the other hand are not seperated and are therefore here treated integrally as line complexes \ion{He}{I} ($\lambda$ 2.112)+\ion{N}{III} ($\lambda$ 2.116) resp. \ion{He}{I} ($\lambda$ 2.162)+\ion{He}{I} ($\lambda$ 2.164)+\ion{H}{I} ($\lambda$ 2.166). All lines resp. line complexes show P Cygni profiles identifying IRS34W as a wind source. We especially note a remarkable similarity between the spectra of IRS34W and the spectrum of the Ofpe/WN9 star HDE 269445 presented in Morris et al. (\cite{Morris}).

The comparison of the spectra shows only slight variations in the equivalent widths (EWs) of all emission lines (see Table \ref{lines}) that are too small to be significant. A direct graphical comparison by overplotting the spectra (see Fig. \ref{spectra}, bottom panel) shows that they appear to be identical within their typical accuracies.

Obviously, the photometric and spectroscopic observations of IRS34W lead to ambivalent results.
On the one hand, the photometric magnitude of the star is highly variable on timescales from months to years. The shape of the light curve and the time scales of the variability rule out the possibility that IRS34W is an eclipsing binary. The amplitude of the variation, up to $\simeq$ 75\% in flux, shows that an eclipsing star
would have to cover at least that fraction of IRS34W's surface on the
sky. The lack of significant variation in the line to
continuum ratios shows that the continuum of this eclipsing star would
have to be small relative to the fraction of the flux of IRS34W that
remains to be seen. In that case, the temperature of the eclipsing star
would be below $\simeq$5000K, and it would have CO absorption lines,
that we would be able to detect even diluted in the continuum of the
blue supergiant. Therefore also a coincidental eclipse by an unrelated star can be excluded.

\begin{table}[t]
\caption{Equivalent widths of the dominant emission lines in the March 1996, April 2003 and August 2004 spectra. Column ``A'' is \ion{He}{I} ($\lambda$ 2.058) , ``B''  \ion{He}{I} ($\lambda$ 2.112) +\ion{N}{III} ($\lambda$ 2.116)  and ``C'' \ion{He}{I} ($\lambda$ 2.162)+\ion{He}{I} ($\lambda$ 2.164)+\ion{H}{I} ($\lambda$ 2.166) . The wavelengths of the lines are given in microns, the equivalent widths are given in \AA.}
\label{lines}
\centering
\begin{tabular}{l c c c}
\hline\hline
 & A & B & C \\
\hline
1996 & 22.1$\pm$1.9 & 3.2$\pm$1.6 & 18.7$\pm$1.9 \\
$2003^{\mathrm{a}}$ & 26.0$\pm$0.6 & 5.4$\pm$0.7 & 18.5$\pm$0.6 \\
$2003^{\mathrm{b}}$ & 25.4$\pm$0.9 & 3.5$\pm$0.8 & 18.5$\pm$0.5 \\
2004 & 25.8$\pm$0.8 & 4.4$\pm$0.9 & 15.6$\pm$0.9 \\
\hline
\end{tabular}

\begin{list}{}{}
\item[$^{\mathrm{a}}$] Spectrum taken from SPIFFI cube reduced to pixel scale, spatial and spectral resolution of 3D data.
\item[$^{\mathrm{b}}$] Spectrum obtained from original SPIFFI cube.
\end{list}

\end{table}

Another possible effect to be examined is the obscuration of IRS34W by interstellar matter.
The central parsec is known to contain several gas
patches of various dimensions, that can be responsible for local
enhancements of the extinction by $\Delta K\simeq 1$ (Paumard et al. \cite{Paumard04}), and could therefore in principle account for the variability of
IRS34W. Such clouds are however several arcseconds wide, and should therefore effect not only IRS34W but also neighbouring stars.
Indeed, the lightcurves of some faint stars ($K \simeq 16$) taken from the original automatically generated NACO photometric data set (Trippe \cite{Trippe}) in a distance up to 0.9'' from IRS34W showed similar, but weaker variations. Therefore we reanalyzed the magnitudes of IRS34W and its neighbouring stars in a set of selected good NACO images, applying additional background flux subtraction and Lucy-Richardson deconvolution instead of Wiener filtering. This analysis confirmed the magnitudes and variations found for IRS34W and made it possible to trace irregularities in some faint stars back to flux spill-over effects from IRS34W to stars located in its seeing halo. This rules out as well a cloud in the line of sight but much closer to
the Earth, as it would be even larger in projection, and cause similar
variability in a large number of stars, that would be obvious in our
data.
So we finally could conclude, that the photometric variability is indeed bound to the star resp. its immediate vicinity (in terms of the given spatial resolution) and not to an external source.

\begin{figure}[t]
\resizebox{\hsize}{!}{\includegraphics{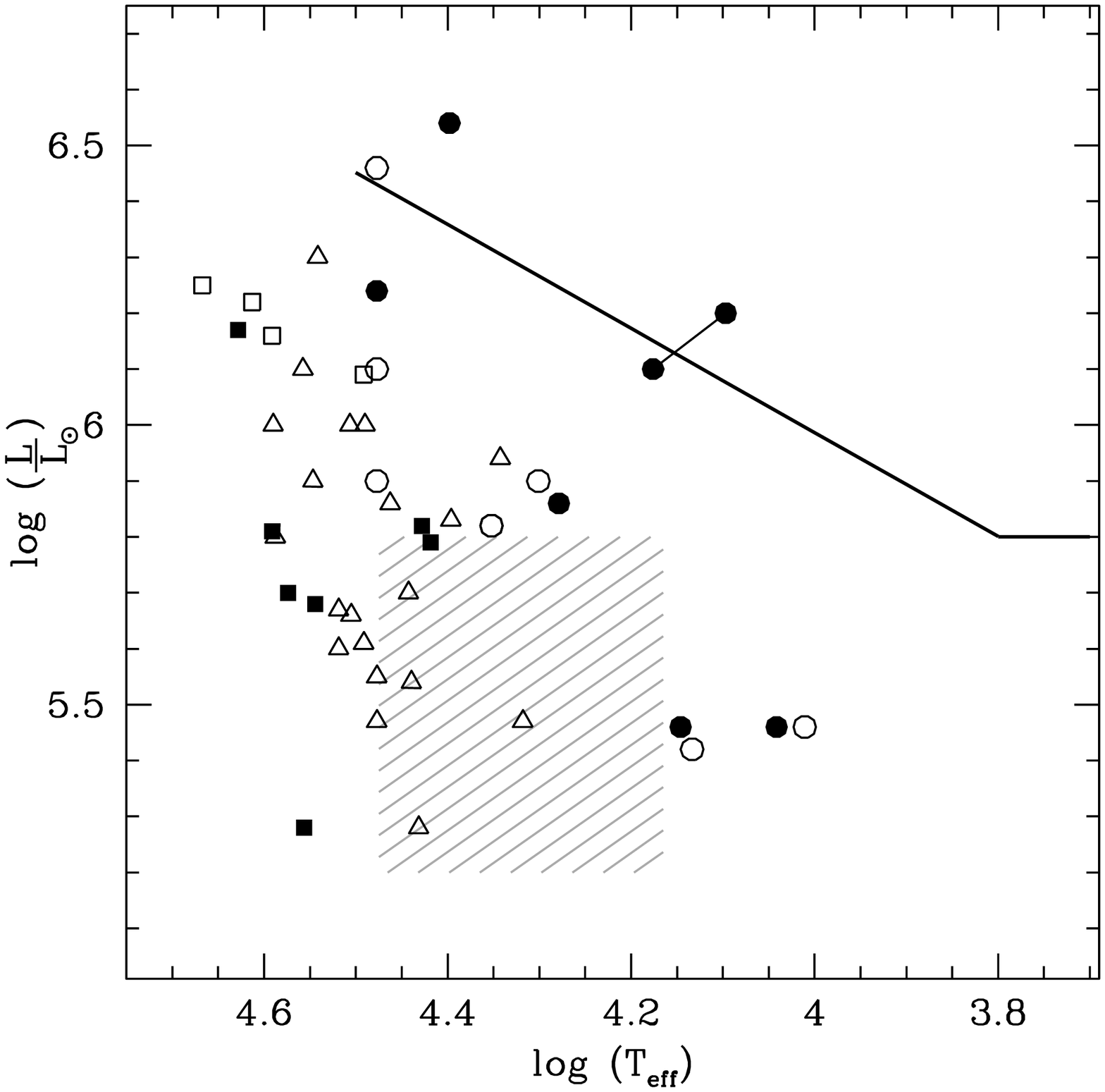}}
\caption{The approximated position of IRS34W (shaded area) in a HRD compared to already known similar stars. Triangles represent Ofpe/WN9 stars (Pasquali et al. \cite{Pasquali}, Crowther et al. \cite{Crowther}, Bianchi et al. \cite{Bianchi}, Bresolin et al. \cite{Bresolin}), circles LBVs (Humphreys \& Davidson \cite{Humphreys94}, Clark et al. \cite{Clark03a}) and squares O supergiants (Bohannan \& Crowther \cite{Bohannan99}, Repolust et al. \cite{Repolust}, Crowther et al. \cite{Crowther02}), filled symbols mark Galactic stars, empty symbols are stars in the Large Magellanic Cloud and NGC 300; the solid line is the Humphreys-Davidson limit. This diagram shows that IRS34W is located between O supergiants and LBVs.}
\label{hrd}
\end{figure}

On the other hand, the absence of significant spectroscopic variability complicates the understanding of the stellar behaviour. As the emission lines we observe in our spectra are sensitive especially to the stellar parameters $T_{\rm eff}$ and $\dot{M}$, significant variations of these parameters should be visible in the spectra. An example for this is the LBV AFGL 2298 that shows photometric variations in $K$ band of around 0.3 magnitudes and variations in the EWs of \ion{He}{I} ($\lambda$ 2.058)  and \ion{H}{I} ($\lambda$ 2.166)  by factors of 3 to 4 within one year (Clark et al. \cite{Clark03a}). Indeed, our observations indicate that at least $T_{\rm eff}$ and $\dot{M}$ remain rather constant while the star gets dimmer or brighter.

Given that we do not see significant changes in the spectral lines,
variations of the wind properties are highly improbable since this would mean a change of the density (by modifications either of maximum velocity at the top of the wind, $v_{\infty}$, or $\dot M$), which would lead to significant observable spectroscopic variations.

Taking this into account, we find two possible interpretations for the photometric variability of IRS34W:

(1) The variability is connected to the physics of the stellar photosphere. Since $T_{\rm eff}$ did not change, this assumption would lead to
the conclusion that the stellar radius (as $L \propto R_{*}^2\cdot T_{\rm eff}^4$) changed by a factor of around 2.
Indeed Dorfi \& Gautschy (\cite{Dorfi}) observe comparable variations in stellar
radii (by a factor up to $\simeq$1.6) in their simulations of pulsations of massive stars, but on timescales of days, not years as in the case of IRS34W.
More interestingly, they also note that the \it raise \rm of a pulsating mode
can be accompanied by a change in stellar radius (a factor of $\simeq$1.3 in
their model) and only a little increase in $T_{\rm eff}$ (2700K on top of
18000K, thus probably not observable for us), on the timescale of two years. Although in this particular case the overall change in luminosity is smaller than the variations we observe ($\Delta K\simeq 0.44$ as calculated from a black-body distribution for the given changes in temperature and radius), it makes it plausible that we may have witnessed a phase
transition between a hydrostatic equilibrium and a pulsating mode.

This explanation is complicated by the fact that a change in stellar radius would change the observed EWs, which scale as $\dot M / R_{*}^{1.5}$. Thus the constancy of the EWs would require a corresponding increase of $\dot M$ parallel to the rise of the radius.

Apart from this rather improbable scenario, it is hard to understand how a
significant change in stellar radius, accompanied by a new hydrostatic
quasi-equilibrium that would last for years, would not be accompanied by
a significant change in $T_{\rm eff}$. 

(2) The photometric variations of IRS34W are not
caused by the photosphere but from above the wind. The colour
of the star, significantly redder than other stars of similar brightness (e.g. our photometric reference star ID185 located 3'' NE of IRS34W with $K = 12.09 \pm 0.08$ and $H-K = 2.26 \pm 0.13$) in the field, also suggests that it is affected by a higher extinction.

This is supported by Clenet et al. (\cite{Clenet}) who derive a colour of $K-L = 2.4$ which is about 1 magnitude redder than the colours of other \ion{He}{I} resp. Ofpe/WN9 stars observed in the field. Additionally, Viehmann et al. (\cite{Viehmann}) find the $H-K$ colours of other Galactic Centre \ion{He}{I} stars to be clearly bluer than what we find for IRS34W.

As we were able to exclude external effects like stars or interstellar clouds, the variability would then be due to variations in the column density of circumstellar material expelled by the star.

Both cases - from which scenario (2) is obviously the more probable one - suggest the identification of IRS34W as a LBV - or close to this stage. It must be noted that a slightly elongated nebular emission
feature is visible around IRS34W on a $H/K/L$ NACO colour map that
traces the dust emission (Genzel et al. \cite{Genzel03a}, Fig. 1 right), and that this feature could as well be an
interstellar structure as circumstellar material from IRS34W that might have been ejected in earlier LBV type eruptions. It is also interesting
to note that the shape of the $K$ band light curve of IRS34W is very similar to the $V$
light curve of the LBV AG Car (see Humphreys \& Davidson \cite{Humphreys94}): a sudden decrease is followed by a plateau and then a slight increase in a time span of around six years. But contrary to IRS34W, in case of AG Car the photometric variability is accompanied by strong spectroscopic variability.

Another interesting point supporting the "dust scenario" is given by
Clark et al. \cite{Clark03b}. Indeed, their mid-IR study of two ring nebulae shows that they
contain stars with properties very similar to candidate LBVs. The K band
spectra clearly identify them as luminous evolved massive stars. They
are also similar to the spectrum of IRS34W, being dominated by H and \ion{He}{I}
emission lines. The surrounding nebulae are explained by
direct ejection of material by the star and/or by interaction of ejected
material with the interstellar medium. Among the two stars, G24.73+0.69
is especially interesting since it is photometrially variable ($\Delta K
\simeq 0.3$) over $\simeq 1.5$ years, but shows only little spectroscopic
variability (see Fig. 10 of Clark et al. \cite{Clark03b}). This similarity favors
the explanation of the observed behaviour of IRS34W by the presence of
circumstellar dust.

In addition, Clark et al. \cite{Clark05} included IRS34W in their list of confirmed
LBVs, although this is partly based on the assumed variation of stellar
and wind parameters that we show is not likely. However, they mention
the possibility of a dusty circumstellar environment to explain the
magnitude variations of IRS34W,  in agreement with the present study.
They note that only LBVs, B[e] supergiants and late WC stars are known
to be dust producers. If dust is indeed the explanation for the
variability of IRS34W, this is then an indication that it is closely
related to Luminous Blue Variables, despite the lack of variation in
stellar properties.

Figure \ref{hrd} shows the approximated position of IRS34W in a Hertzsprung-Russell diagram. The luminosity
was estimated as follows: $A_K$ was derived from the $V$ band extinction map
of Scoville et al. (\cite{Scoville}) and the extinction law of Moneti et al.
(\cite{Moneti}), and was used together with the observed $K$ magnitude to estimate $M_K$. Then, bolometric corrections computed from our preliminary modelling of Ofpe/WN9 stars covering a wide range of physical parameters ($L$, $T_{\rm eff}$,
$\dot M$) were used to finally derive $L$. The extension of the area takes into account
uncertainties in the extinction and the bolometric corrections. As for the
effective temperature, we only show a range of values for which, within
the $L$ range, \ion{He}{I} ($\lambda$ 2.058) shows a P-Cygni profile in our preliminary
models. So far this is a rather rough estimate, but gives some
interesting indications on the evolutionary status of this star. Compared to already known Ofpe/WN9 stars, LBVs and O supergiants, IRS34W is located between the O supergiant and LBV stages (like other known Ofpe/WN9 stars, see Fig. \ref{hrd}) and therefore appears to be an object in transition between these two phases.

The estimate of the luminosity of IRS34W presented above shows that it may be
intrinsically less luminous than the other \ion{He}{I} stars (see Najarro et al. \cite{Najarro}). As discussed above, it is, regardless of the variations reported
here, also redder. Given this and the fact that dust formation
is the most likely explanation for the observed variability, we can
imagine the following scenario: due to its lower luminosity, IRS34W is
able to experience dust formation in its atmosphere. Such episodes of
dust formation makes the star fainter and redder;
at the end of each episode, dust is destroyed, but only partially,
leading to a long-term accumulation of dust around the star so that on
average, IRS34W is redder (and also even fainter) than the other \ion{He}{I}
stars. The key point of this scenario is the lower intrinsic luminosity
of the star previous to any dust formation episode.

This is very
speculative however since the formation of dust in atmospheres of hot
stars is still poorly understood, although Cherchneff \& Tielens (\cite{Cherchneff}) argue that special
geometries such as equatorial disks can lead to the high densities
necessary for dust formation; other properties such as a thick wind or inhomogeneities in the atmosphere may also be necessary. In the other \ion{He}{I} stars of the Galactic
Center, the luminosity may be too high, so that dust formation is
inhibited by the too strong radiation. In IRS34W, the photon flux may be
low enough to allow dust formation.

\section{Summary}

In this article we reported the results of time-resolved photometry and spectroscopy of the Galactic Centre star IRS34W and possible interpretations, which can be summarized as follows:

\noindent
1. The photometric magnitude of IRS34W is highly variable on timescales from months to years with an overall variation of $\Delta K \simeq 1.5$.

\noindent
2. Variations of the stellar colour $H-K$ track the photometric variability. When the brightness decreases, the colour gets redder, when the brightness increases, the colour gets bluer.

\noindent
3. The spectra of the star do not show any significant variations over a time span of 8 years.

\noindent
4. The most probable reason for the stellar behaviour is an obscuration by circumstellar material ejected by the star.

\noindent
5. IRS34W is a star in transition between the O supergiant and LBV phases, only the lack of spectroscopic variability prevents us from safely identifying it as a LBV.

IRS34W and five additional stars, which are very similar to IRS34W, have been classified as LBV candidates by Paumard et al. (\cite{Paumard01}). Among these additional stars only one, IRS16SW, shows strong, periodic variability (Ott et al. \cite{Ott99}). Spectrophotometric monitoring of these stars is currently under way; detailed quantitative analysis and modelling of this group of stars, including IRS34W, should shed more light on the physics of this class of objects.

\begin{acknowledgements}

We thank the referee, F. Najarro, for helpful suggestions and
comments which helped to improve the quality of the paper. FM
acknowledges support from the Alexander von Humboldt Foundation.

\end{acknowledgements}

\end{document}